\documentclass[aps,prl,twocolumn,superscriptaddress,longbibliography]{revtex4-2}
\usepackage{graphicx}
\usepackage{amssymb,amsmath}
\usepackage{bm}
\usepackage{dcolumn}
\usepackage{float}
\usepackage[OT1]{fontenc} 
\usepackage{url}
\usepackage{mathrsfs}
\usepackage{slashed,comment}
\usepackage{color}
\usepackage{verbatim}
\usepackage{enumitem}
\usepackage{soul,physics}
\usepackage[driverfallback=dvipdfm]{hyperref}
\hypersetup{pdfpagemode=FullScreen,colorlinks=true,breaklinks,urlcolor=blue,linkcolor=blue,citecolor=blue}

\usepackage{subfigure}
\usepackage{amssymb}
\usepackage{bm}
\usepackage{graphicx}
\usepackage{amsmath,amssymb}

\usepackage{pifont}

\begin{document}
 
  \title{Unitary Designs from Two Chaotic Hamiltonians and a Random Pauli Operation  }

  \author{Ning Sun}
  \affiliation{State Key Laboratory of Surface Physics \& Department of Physics, Fudan University, Shanghai, 200438, China}

  \author{Pengfei Zhang}
  \thanks{PengfeiZhang.physics@gmail.com}
  \affiliation{State Key Laboratory of Surface Physics \& Department of Physics, Fudan University, Shanghai, 200438, China}
  \affiliation{Hefei National Laboratory, Hefei 230088, China}

  \date{\today}

  \begin{abstract}
   The realization of unitary designs is of fundamental interest in quantum science and typically requires the ability to implement structured quantum circuits. Recent developments have explored the possibility of generating unitary designs using only a small number of quantum quenches, in which the evolution during each interval is governed by a static Hamiltonian. In particular, it has been established that at least three chaotic Hamiltonians are required when only Hamiltonian evolutions are employed. In this work, we propose the emergence of unitary designs in the temporal ensemble of qubit systems evolved under two distinct chaotic Hamiltonians for sufficiently long times, supplemented by an intermediate random Pauli operation inserted between them. This result follows from the universal Pauli spectrum of chaotic Hamiltonians, a central concept in the study of non-stabilizerness. Our theoretical predictions are verified numerically using explicit examples, including Gaussian unitary ensemble Hamiltonians and random spin models. We further investigate finite-time and finite-size corrections to the protocol. Our results provide new insights into the dynamical generation of quantum randomness and offer a new route toward realizing unitary designs in chaotic systems.
  \end{abstract}
    
  \maketitle

  \emph{ \color{blue}Introduction.---} Understanding how the complexity of evolution operators grows in chaotic many-body systems is a central topic in modern quantum many-body physics, as it is intimately linked to information scrambling~\cite{Sekino:2008he,Hayden:2007cs,kitaev2014talk,Shenker:2013pqa,Shenker:2014cwa,Maldacena:2015waa,Roberts:2014isa} and quantum thermalization~\cite{1999JPhA...32.1163S,PhysRevA.43.2046,Kaufman:2016mif}. A key concept is unitary $k$-designs~\cite{DiVincenzo:2001lru,PhysRevA.80.012304,Ambainis:2007ura,Emerson:2005oof,Scott:2008ruh,Roberts:2016hpo,Cotler:2017jue}, which refer to ensembles of unitary operators whose first $k$ moments coincide with those of the Haar-random unitary ensemble. For example, random Pauli operators form a unitary 1-design~\cite{Roberts:2016hpo}, leading to thermalization to an infinite-temperature ensemble. On the other hand, random Clifford operations in qubit systems form a unitary 3-design~\cite{Webb:2015vwz,PhysRevA.96.062336} and result in the saturation of out-of-time-order correlators (OTOCs)~\cite{Larkin1969QuasiclassicalMI}. The recent study on higher-order OTOCs~\cite{Abanin:2025rbz,Vardhan:2025rky,Guo:2025hwe,Liu:2026nnw} is also connected to higher-order designs. In addition, unitary designs have broad applications in quantum computation: they can be used for randomized measurements~\cite{Qi:2019rpi,Huang:2020tih,Elben:2022jvo} and serve as a resource for demonstrating quantum advantage~\cite{Arute2019QuantumSU,Zhong2020QuantumCA,Madsen2022QuantumCA}.

  \begin{figure}[t]
    \centering
    \includegraphics[width=0.75\linewidth]{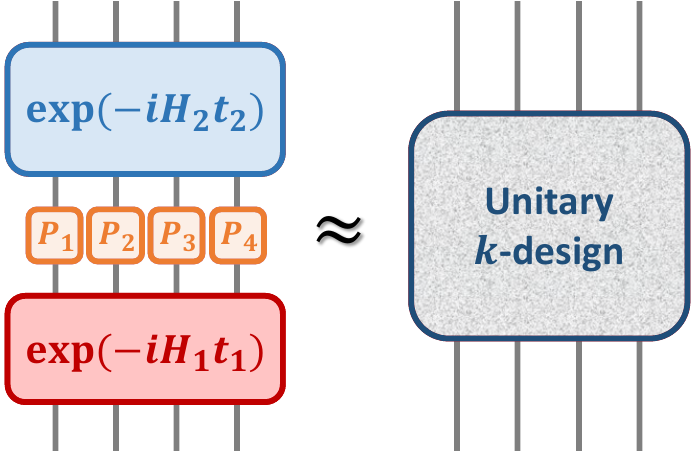}
    \caption{The schematic illustration of our main results. We consider the temporal ensemble generated by evolution under two fixed chaotic Hamiltonians $H_1$ and $H_2$, with a random Pauli operation inserted between them. We show that this protocol enables the emergence of unitary $k$-designs at sufficiently long times in the thermodynamic limit.  }
    \label{fig:schematics}
  \end{figure}

  As a consequence, generating unitary designs in realistic setups has become an important question. In particular, it has been established that the ability to implement structured quantum circuits enables the generation of unitary designs~\cite{Emerson:2003www,Dankert:2005jqr,Gross:2007xgw,Ambainis:2007ura,Roy:2008ngh,PhysRevLett.116.170502,Brandao:2012zoj,PRXQuantum.2.030316,Schuster:2024ajb,Hou:2025bau,Zhang:2025dhg}, with an optimized depth that scales logarithmically with system size~\cite{Schuster:2024ajb}. This makes the circuit-based approach well suited to quantum computing platforms. In contrast, quantum simulation platforms typically lack the ability to implement complex quantum circuits. Therefore, understanding how to generate unitary designs through Hamiltonian evolution has likewise attracted significant attention~\cite{PhysRevA.97.023604,PRXQuantum.2.030316,Onorati:2016met,Jian:2022pvj,Tiutiakina:2023ilu,Guo:2024zmr,Zhou:2025noh}. Recently, Zhou et al. have shown that unitary designs can be realized either through a single quantum quench separating evolutions governed by two random Hamiltonians~\cite{Zhou:2025noh}, or within the temporal ensemble generated by evolution under three fixed chaotic Hamiltonians~\cite{Zhou:2026ubz}. To our knowledge, these constitute the simplest schemes for realizing unitary designs when only Hamiltonian evolutions are employed.

  In this work, we take one step back: in addition to Hamiltonian evolution, we allow random Pauli operations, which correspond to applying a random Pauli operator to each qubit. Such operations can be implemented by applying control fields to specific subsystems and are therefore feasible on quantum simulation platforms. With this new ingredient, we show that two chaotic Hamiltonians are sufficient to generate unitary designs in the temporal ensemble, as illustrated in Fig.~\ref{fig:schematics}. Interestingly, this mechanism is closely related to the universal property of the Pauli spectrum in generic chaotic systems~\cite{Beverland:2019jej,Turkeshi:2023lqu}, which has recently been proposed in the context of quantum magic~\cite{PhysRevA.71.022316,PhysRevA.86.052329,Veitch:2012ttw,Emerson:2013zse}. We explicitly verify our protocol using concrete Hamiltonians sampled from the Gaussian unitary ensemble (GUE) and from random spin models. Our results provide a new route toward generating unitary designs in many-body systems and shed new light on the interplay between quantum many-body chaos and quantum magic.

  \emph{ \color{blue}Setup.---} We now describe our scheme, illustrated in Fig.~\ref{fig:schematics}. We focus on systems consisting of $N$ qubits. The system is first evolved under a chaotic Hamiltonian $H_1$ for a time $t_1$, where $t_1$ is sampled uniformly from the interval $[0, T]$. We then apply a random Pauli operation with unitary operator $P = P_1 P_2 \cdots P_N$, where $P_i \in \{I, X, Y, Z\}$ denotes a single-qubit Pauli operator acting on site $i$. For now, we assume that all Pauli operators are chosen with equal probability. The consequences of modifying the Pauli ensemble will be discussed later. Finally, we apply a second Hamiltonian evolution governed by a distinct Hamiltonian $H_2$ for a time $t_2$, where $t_2$ is sampled from the same distribution as $t_1$. The resulting ensemble of unitary operators is given by
  \begin{equation}\label{eq:defE}
   \mathcal{E}=\big\{e^{-iH_2t_2}Pe^{-iH_1t_1},\big|t_1,t_2\in[0,T], P\big\}.
  \end{equation}

  We aim to study whether the ensemble $\mathcal{E}$ forms a unitary $k$-design in the thermodynamic limit $N \rightarrow \infty$. This is typically quantified by the $k$-th frame potential $F^{(k)}_{\mathcal{E}}$~\cite{Scott:2008ruh,Roberts:2016hpo,Cotler:2017jue}, defined as
  \begin{equation}\label{eq:FPdef}
  F^{(k)}_{\mathcal{E}}\equiv \mathbb{E}_{U,V\in \mathcal{E}}~\left|\text{tr}[V^\dagger U]\right|^{2k},\ \ \ k\in\mathbb{N^+}.
  \end{equation}
  It has been shown that the Haar-random unitary ensemble achieves the minimal $k$-th frame potential among all ensembles of unitary operators, with $F^{(k)}_{\text{Haar}} = k!$~\cite{Scott:2008ruh}. Therefore, an ensemble $\mathcal{E}$ forms a unitary $k$-design if and only if $F^{(k)}_{\mathcal{E}} = k!$. The frame potential can also be related to the R\'enyi entropy of EPR states when one member of each pair is subjected to noisy unitary evolution drawn from the ensemble $\mathcal{E}$~\cite{Wang:2023vkq}. 

  \emph{ \color{blue}General arguement.---} We proceed to calculate the frame potential for our protocol $\mathcal{E}$, defined in Eq.~\eqref{eq:defE}. Random Pauli operations with a uniform distribution already form an exact 1-design, and this property remains valid after composition with Hamiltonian evolutions~\cite{Roberts:2016hpo}. Therefore, it suffices to analyze the case $k \geq 2$. Now, we focus on $k = 2$ as a concrete example. We represent the eigenstates and eigenvalues of the two Hamiltonians as $H_1 |E_i\rangle = E_i |E_i\rangle$ and $H_2 |\epsilon_a\rangle = \epsilon_a |\epsilon_a\rangle$, where $i, a \in \{1, 2, \dots, D\}$ with the Hilbert space dimension $D = 2^N$. This allows us to expand 
  \begin{equation}\label{eq:single}
  \text{tr}[V^\dagger U]=\sum_{i,a}e^{-i\delta t_1 E_{i}-i\delta{t_2}\epsilon_{a}}\langle \epsilon_{a}|P|E_{i}\rangle \langle E_{i}|\tilde{P}|\epsilon_{a}\rangle.
  \end{equation}
  Here, $P$ and $\tilde{P}$ denote the Pauli operators associated with $U$ and $V$, respectively. We have also introduced $\delta t_r = t_r - t_r'$ with $r\in \{1,2\}$, representing the differences between the random evolution times $t_r$ and $t_r'$ for $U$ and $V$. To calculate $F_\mathcal{E}^{(2)}$, we introduce four replicas of Eq.~\eqref{eq:single} labeled by $n\in\{1,2,3,4\}$. The result reads
  \begin{equation}
  \begin{aligned}
  F_\mathcal{E}^{(2)}=\sum_{i_n,a_n} I(E_{i_n},&\epsilon_{a_n})~P_{a_1i_1} P_{a_2i_2}^*P_{a_3i_3} P_{a_4i_4}^*\\
  \times&\tilde{P}_{i_1a_1} \tilde{P}_{i_2a_2}^*\tilde{P}_{i_3a_3} \tilde{P}_{i_4a_4}^*.
  \end{aligned}
  \end{equation}
  Here, we keep the average over $P$ and $P'$ implicit and introduce $P_{ai}=\langle \epsilon_{a}|P|E_{i}\rangle$ and $\tilde{P}_{ia}=\langle E_{i}|\tilde{P}|\epsilon_{a}\rangle$ for brevity. The time-filter function $I(E_{i_n},\epsilon_{a_n})$ reads
  \begin{equation}
  I(E_{i_n},\epsilon_{a_n})\equiv \mathbb{E}_{\delta t_r}e^{i\sum_{n=1}^4 (-1)^n (E_{i_n}\delta t_1 +\epsilon_{a_n}\delta t_2)},
  \end{equation}
  where $(-1)^n$ arises from the complex conjugation in the replicas involving $U^\dagger V$. When the total evolution time $T \rightarrow \infty$, time averaging over $\delta t_1$ imposes the constraint $E_{i_1} + E_{i_3} = E_{i_2} + E_{i_4}$. In the absence of two-level resonances in chaotic systems, the only solutions are $(i_1=i_2, i_3=i_4)$ or $(i_1=i_4, i_3=i_2)$. Similar relations holds for $a_n$. This is referred to as the perfect time-filter limit in~\cite{Zhou:2026ubz}. We find
  \begin{equation}\label{eq:F2}
  \begin{aligned}
  \frac{F_\mathcal{E}^{(2)}}{2}=&\sum_{i_n,a_n}P_{a_1i_1} P_{a_1i_1}^*P_{a_3i_3} P_{a_3i_3}^*\tilde{P}_{i_1a_1} \tilde{P}_{i_1a_1}^*\tilde{P}_{i_3a_3} \tilde{P}_{i_3a_3}^*\\& + P_{a_1i_1} P_{a_1i_3}^*P_{a_3i_3} P_{a_3i_1}^*\tilde{P}_{i_1a_1} \tilde{P}_{i_1a_3}^*\tilde{P}_{i_3a_3} \tilde{P}_{i_3a_1}^*.
  \end{aligned}
  \end{equation}
  The first and second lines represent the contributions in which the solutions for the $i$ and $a$ either share the same pairing pattern or follow different pairing patterns.

  The evaluation of Eq.~\eqref{eq:F2} therefore requires analyzing the matrix elements $P_{ai} = \langle \epsilon_a | P | E_i \rangle$ for typical Pauli operators $P$. To gain some intuition, we consider the scenario in which both $H_1$ and $H_2$ are sampled from the GUE~\cite{2004math.ph..12017F}. In this case, the corresponding eigenstates are Haar-random states, which can be written as $|E_i\rangle = W_1 |i\rangle$ and $|\epsilon_a\rangle = W_2 |a\rangle$, where $W_1$ and $W_2$ are typical Haar-random unitaries. As a consequence, $P_{ai}=\langle a|W_2^\dagger PW_1|i\rangle$ is also Haar random. We further express
  \begin{equation}\label{eq:P'}
  \tilde{P}_{ia}=\langle i|W_1^\dagger \tilde{P} W_2|a\rangle=\sum_{c}P_{ci}^* \langle c|W_2^\dagger P_rW_2|a\rangle, 
  \end{equation} 
  with $P_r\equiv P\tilde{P}$. Next, we neglect sample fluctuations and perform the Haar average over $P_{ai}$. Standard Weingarten calculus~\cite{Roberts:2016hpo} in the limit $D \rightarrow \infty$ implies that matrix elements $P_{ai}$ with different indices behave as independent Gaussian variables with variance $1/D$. We assume this remains valid for general chaotic Hamiltonians.

  To extract the dominant contribution to Eq.~\eqref{eq:F2}, we should avoid introducing additional constraints on the indices $i_n$ and $a_n$, as such constraints would render the contribution subleading in $1/D$. We begin by considering the special case $P = \tilde{P}$, which occurs with probability $p_0 = 4^{-N}$ for Pauli operations drawn from a uniform distribution. In this case, the Haar average takes the same form as in the two-step protocol without Pauli operations, which has already been analyzed in~\cite{Zhou:2026ubz}. In both lines of Eq.~\eqref{eq:F2}, Wick’s theorem allows contractions between $P_{a_ni_n}$ and $\tilde{P}_{i_na_n}=P_{a_ni_n}^*$. In addition, the first line permits an extra contraction pattern between $P_{a_ni_n}$ and $P_{a_ni_n}^*$. Putting all these contributions together, we find that the resulting contribution to $F^{(2)}_{\mathcal{E}}$is $6p_0$, consistent with Ref.~\cite{Zhou:2026ubz}. Next, we consider the contribution from $P \neq P'$. The key difference is that, if we attempt to contract $P_{a_n i_n}$ with $\tilde{P}_{i_n a_n}$, Eq.~\eqref{eq:P'} predicts
  \begin{equation}
  \overline{P_{a_n i_n}\tilde{P}_{i_n a_n}}=\langle \epsilon_a| P_r|\epsilon_a\rangle/D.
  \end{equation}
  The collection of $c(P_r)=\langle \epsilon_a| P_r|\epsilon_a\rangle$ is known as the Pauli spectrum~\cite{Beverland:2019jej,Turkeshi:2023lqu}, which provide detailed information of the many-body state $|\epsilon_a\rangle$. It has been established that for chaotic systems $c(P_r)$ satisfies a real Gaussian distribution with a zero mean and a variance of $1/(D+1)$~\cite{Turkeshi:2023lqu}. This leads to an additional suppression, causing the corresponding contraction to vanish in the thermodynamic limit. Therefore, the only remaining contribution comes from the contraction between $P_{a_n i_n}$ and $P_{a_n i_n}^*$ in the first line. Putting all ingredients together, we find 
  \begin{equation}
  F^{(2)}_{\mathcal{E}}=6p_0+2(1-p_0)~\xrightarrow{N\rightarrow\infty}~2.
  \end{equation}

  Finally, we briefly describe the generalization of the above analysis to arbitrary $k$. Similar to Eq.~\eqref{eq:F2}, in the perfect-filter limit, the calculation of $F^{(k)}$ requires analyzing the $2k$-th moments of $P_{a_n i_n}$ and $\tilde{P}_{i_n a_n}$, where the indices $i_n$ and $a_n$ can follow different pairing patterns. However, for $P \neq \tilde{P}$, only the terms with the same pairing pattern contribute in the thermodynamic limit. This reduces the double sum over the permutation group to a single sum, resulting in a contribution of $(1 - p_0) k!$. Adding the contribution from the special case with $P = \tilde{P}$, denoted as $F^{(k)}_{\text{2SP}}$~\cite{Zhou:2026ubz}, we obtain
  \begin{equation}\label{eq:resFk}
  F^{(k)}_{\mathcal{E}}=p_0F^{(k)}_{\text{2SP}}+(1 - p_0)k! ~\xrightarrow{N\rightarrow\infty}~k!.
  \end{equation}
  Here, we have used the fact that $F^{(k)}_{\text{2SP}}$ is independent of the system size $N$. This demonstrates the emergence of the unitary $k$-design in our protocol for sufficiently long evolution times in the thermodynamic limit. We further note that, for a Hamiltonian ensemble consisting of a single evolution followed by random Pauli operations, the contribution from $P = \tilde{P}$ already leads to $F_{\mathcal{E}}^{(k)} \geq p_0 k! D^{k}$, which diverges for $k > 2$~\cite{Roberts:2016hpo}. Consequently, our protocol represents the simplest scheme that combines Hamiltonian evolutions with random Pauli operations to generate unitary designs.

  \begin{figure}[t]
    \centering
    \includegraphics[width=0.99\linewidth]{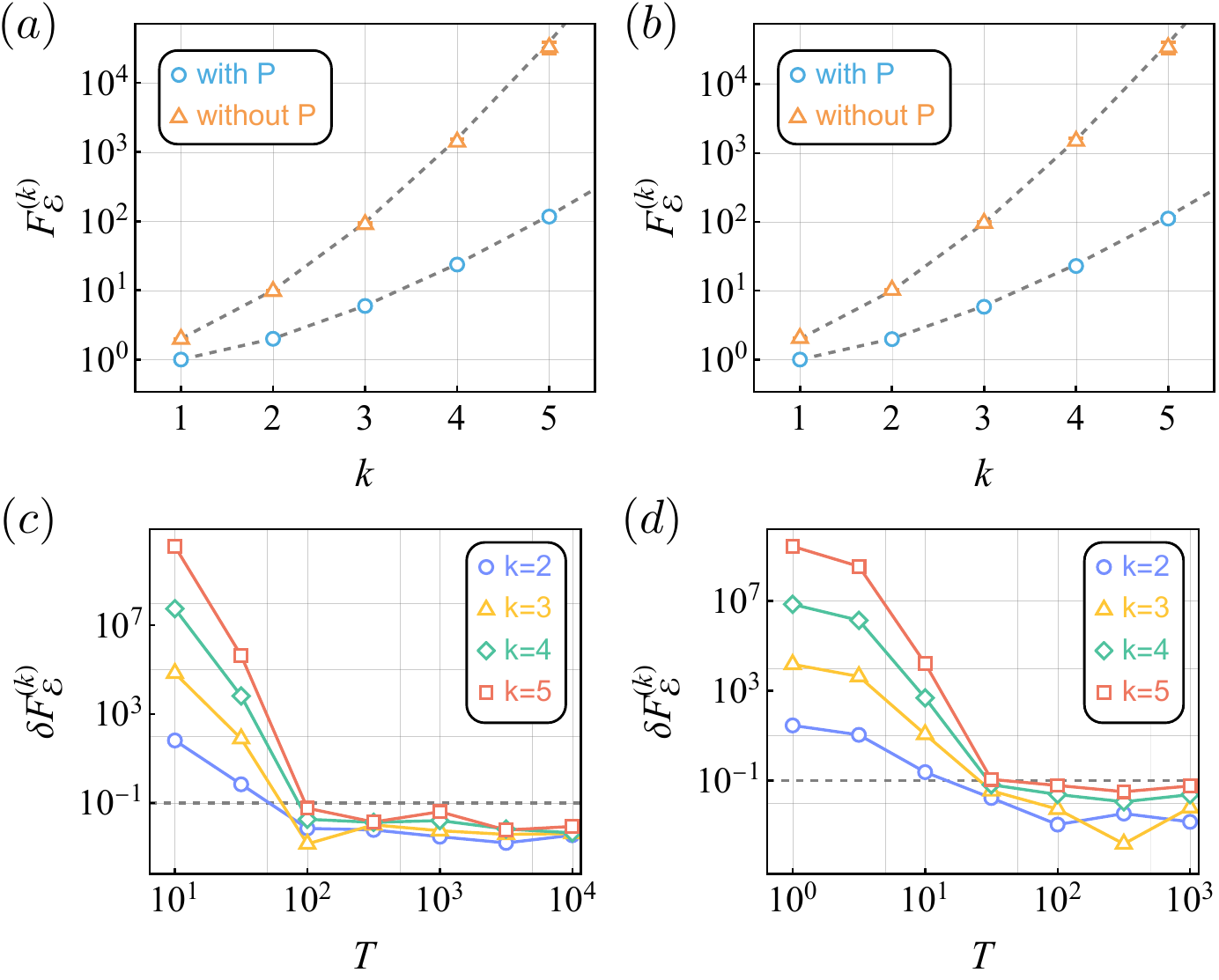}
    \caption{The numerical results of our protocol. The Hamiltonians are drawn from either the GUE (a,c) or random spin models (b,d), for a system size of $N=7$. The results are obtained by generating 400 random evolution operators and computing their overlaps according to Eq.~\eqref{eq:FPdef}. In panels (a-b), we set $T = 10^6$ to ensure sufficiently long evolution times, and we compare protocols with and without the Pauli operations. The dashed lines correspond to either $F_{\text{2SP}}^{(k)}$ or $k!$. The error bars, which reflect statistical fluctuations in the evaluation of the expectation, are smaller than the marker size for most data points. In panels (c-d), the dashed lines indicate the error threshold $\epsilon = 10^{-1}$.  }
    \label{fig:Num1}
  \end{figure}

  \emph{ \color{blue}Numerical demonstration.---} We now provide numerical demonstrations of our theoretical predictions. We consider two different models for the chaotic Hamiltonians: the GUE and random spin models. For the GUE Hamiltonians, the matrix elements $H_{ij}$ are sampled as independent Gaussian variables with variance $1/2D$, subject to the constraint of Hermiticity. The resulting many-body spectrum follows a semicircular distribution with a radius of $\sqrt{2}$~\cite{2004math.ph..12017F}. For random spin models, the Hamiltonian reads
  \begin{equation}
  H=\sum_{i<j,PP'}J_{ij}^{PP'} P_iP'_j+\sum_{i,P}h_i^P P_i.
  \end{equation}
  Here, random couplings $J_{ij}^{PP'}$ and and random fields $h_i^P $ are independently sampled from $[-1/\sqrt{N},1/\sqrt{N}]$ and $[-1,1]$, respectively, which ensures that the eigenstate energies remain extensive. Analogous models can be realized on quantum simulation platforms~\cite{Wei:2021wko,Geier:2021uxg,Garttner:2016mqj,Joshi:2021fno,Zu:2021irm,Morgan:2008hpm}. In the numerics, we first sample $H_1$ and $H_2$ from the chosen Hamiltonian ensemble and then keep them fixed when evaluating the frame potentials for various $k$.

  The numerical results are presented in Fig.~\ref{fig:Num1} for a moderate system size of $N=7$, corresponding to a Hilbert space dimension of $D = 128$. We find that both GUE Hamiltonians and random spin models enable the generation of unitary designs with good accuracy using our protocol with two fixed Hamiltonians and an intermediate random Pauli operation. In contrast, the protocol without random Pauli operators, corresponding to $p_0 = 1$ in Eq.~\eqref{eq:resFk}, deviates significantly from the unitary designs, consistent with Ref. \cite{Zhou:2026ubz}. 

  \begin{figure}[t]
    \centering
    \includegraphics[width=0.99\linewidth]{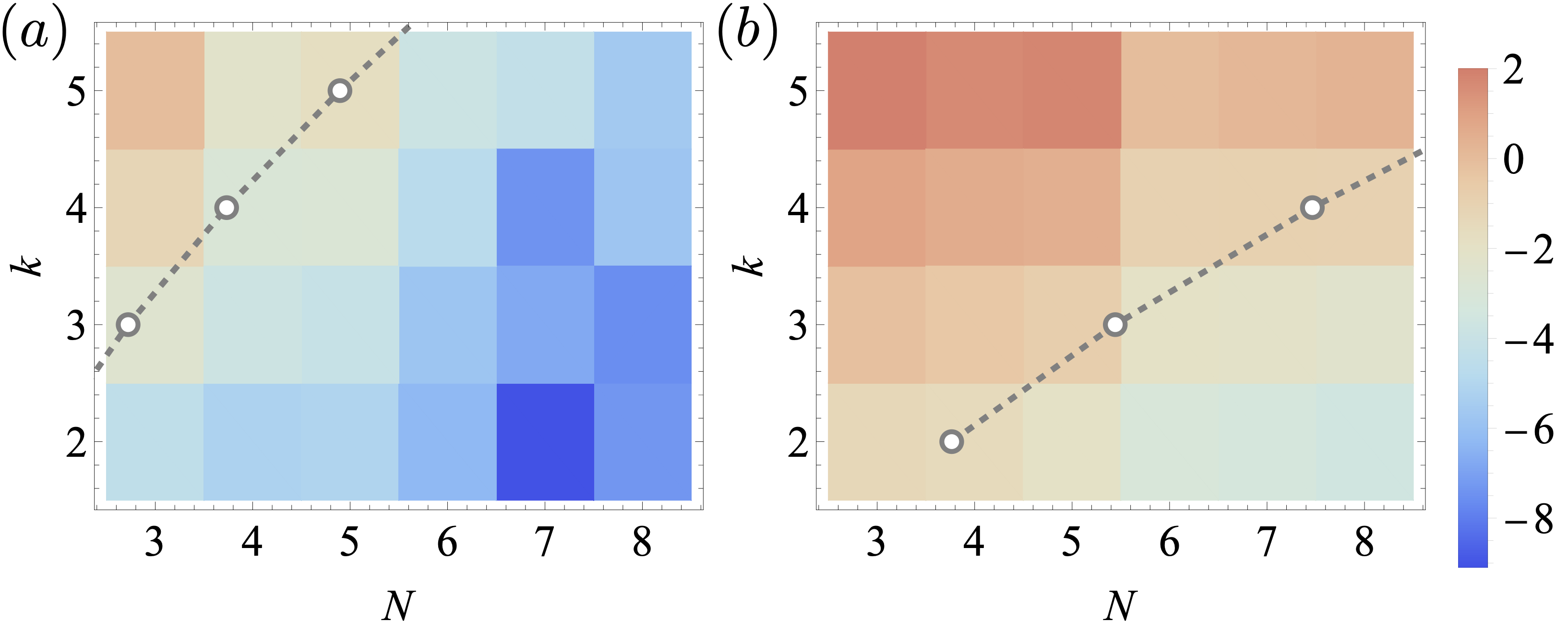}
    \caption{The numerical results for the finite-size corrections of our protocol are shown. We present density plots of $\ln \delta F^{(k)}_\mathcal{E}$ for various values of $k$ and $N$, with $T = 10^6$, for GUE Hamiltonians. For each system size $N$, we generate 400 random evolutions. In panel (a), the Pauli operation is sampled uniformly from all Pauli strings, while in panel (b), it is restricted to operators containing only $I$ or $Z$ on each qubit. The white circles indicate the predicted critical system sizes described in the main text with $\eta=e^{-1}$, and the dashed lines serve as guides to the eye. }
    \label{fig:Num2}
  \end{figure}

  Next, we consider the correction from finite evolution time $T$. The time-filter function at finite time allows for unpaired indices, with a typical suppression of $\text{sinc}(\delta E T)^2$~\cite{Sakurai:2011zz}. Here, the difference between energy levels $\delta E$ can be estimated by the level spacing of many-body spectrum. The additional factor of $D$ arising from the introduction of extra indices is canceled by the suppression resulting from unpaired indices being incompatible with the dominant contraction pattern. This behavior is analogous to that observed in the three-step protocol of Ref.~\cite{Zhou:2026ubz}. Therefore, we expect the deviation of the frame potential to scale as $1/(\delta E T)^2$ at long times. For the GUE Hamiltonian, the energy levels span a finite window, and we therefore expect the critical time required to realize unitary designs to scale as $T_c \sim D =128$. In contrast, the random spin model has an extensive spectrum width, leading to $T_c \sim D/N \approx 18$. In Fig.~\ref{fig:Num1}(c-d), we numerically compute the normalized deviation of the frame potential for different evolution times $T$, defined as $\delta F^{(k)}_{\mathcal{E}} = |F^{(k)}_{\mathcal{E}} - k!|/k!$. The numerical results are consistent with the above analysis.

  We further discuss the consequences of modifying the ensemble of Pauli operations. As an example, we consider restricting the ensemble to Pauli strings that contain only $I$ or $Z$ operators on each site. Physically, this corresponds to applying magnetic fields along a fixed direction for a subset of qubits, which is less challenging for experimental realizations. By examining the derivation in the previous section, it is evident that this modification only changes $p_0$ to $2^{-N}$ and does not affect the result in the thermodynamic limit. Nevertheless, for moderate system sizes accessible to numerical simulations, this modification significantly enhances finite-size effects. The enhancement originates from the rapid growth of $F^{(k)}_{\text{2SP}} \approx e (k!)^2$ as a function of $k$~\footnote{We expect that other finite-size corrections are subleading compared with the contribution arising from finite $p_0$, since they are not enhanced by a factor of $(k!)^2$, as confirmed by the numerical results.}. Requiring the corresponding contribution is suppressed by a factor of $\eta$, we obtain the condition
  \begin{equation}
   p_0(N) \lesssim \eta k!/F^{(k)}_{\text{2SP}} \sim \eta/(e k)!.
  \end{equation}
  This condition determines a critical system size $N$ for a given ensemble of Pauli operations at each value of $k$. We compare this prediction with numerical simulations for above ensembles of Pauli operations, as shown in Fig.~\ref{fig:Num2}(a-b). Here we take $\eta=1/e$ and fix $T=10^6$. The critical system size, represented by the white circles, clearly separates the regime in which $\delta F^{(k)}_{\mathcal{E}}$ is $O(1)$ from the regime in which it scales as $O(1/N)$, consistent with our theoretical estimation. Finally, we may also consider ensembles of Pauli matrices with spatial locality. For example, we sample $l\in\{0,1,\cdots N\}$ uniformly and apply $Z$ operators to sites $i\in [1,l]$. In the the limit of $N\rightarrow \infty$, unitary designs also emerge. However, since $p_0=1/(N+1)$, achieving this requires a system size that grows factorially with $k$. 
  
  \emph{ \color{blue}Discussion.---} In this work, we demonstrate that, although strict Hamiltonian evolution typically requires at least three distinct chaotic Hamiltonians to generate unitary designs, inserting a random Pauli operation reduces this number to two. By analyzing the limits of long evolution time and large system size, we show that the frame potential approaches the Haar-random value. This result relies crucially on the universal operator spectrum of chaotic systems, which suppresses most contractions between random matrix elements. Our numerical simulations for both GUE and random spin models confirm these predictions, highlighting the robustness of the protocol across different chaotic systems. Furthermore, we analyze finite-time corrections, which scale inversely with the square of the evolution time, as well as finite-size effects that originate from the coincidence of Pauli operators between two quantum trajectories.

  We conclude with several remarks on future directions. First, our analysis focuses on inserting a random Pauli operation between two chaotic Hamiltonian evolutions. It is natural to expect that similar protocols, where Pauli operations are replaced by evolutions under integrable Hamiltonians, should also work. Second, while our discussions assume long evolution times, the inevitable effects of decoherence have not been taken into account, which are crucial for practical experimental realizations. Finally, it is interesting to ask what the simplest protocol would be if we allow Hamiltonian evolutions combined with arbitrary Clifford operations. We leave these questions for future investigation.

\vspace{5pt}
\textit{Acknowledgement.} We sincerely thank Yi-Neng Zhou and Tian-Gang Zhou for sharing their manuscript on related topics, which inspired and motivated this study. This project is supported by the Shanghai Rising-Star Program under grant number 24QA2700300, the NSFC under grant 12374477, the Quantum Science and Technology-National Science and Technology Major Project 2024ZD0300101, 2025ZD0300100, and 2025ZD0300101, the Shanghai Municipal Science and Technology Major Project Grant No. 24DP2600100, and Co-research Program under Grant No. 25LZ2601000. PZ is also supported by the Xuemin Institute of Advanced Studies at Fudan University.

\bibliography{ref.bib}

\end{document}